# Framework for Clique-based Fusion of Graph Streams in Multi-function System Testing

Mark Sh. Levin

*Abstract*—The paper describes a framework for multi-function system testing. Multi-function system testing is considered as fusion (or revelation) of clique-like structures. The following sets are considered: (i) subsystems (system parts or units/components/modules), (ii) system functions and a subset of system components for each system function, and (iii) function clusters (some groups of system functions which are used jointly). Test procedures (as units testing) are used for each subsystem. The procedures lead to an ordinal result (states, colors) for each component, e.g., $[1, 2, 3, 4]$ (where $1$ corresponds to 'out of service', $2$ corresponds to 'major faults', $3$ corresponds to 'minor faults', $4$ corresponds to 'trouble free service'). Thus, for each system function a graph over corresponding system components is examined while taking into account ordinal estimates/colors of the components. Further, an integrated graph (i.e., colored graph) for each function cluster is considered (this graph integrates the graphs for corresponding system functions). For the integrated graph (for each function cluster) structure revelation problems are under examination (revelation of some subgraphs which can lead to system faults): (1) revelation of clique and quasi-clique (by vertices at level 1, 2, etc.; by edges/interconnection existence) and (2) dynamical problems (when vertex colors are functions of time) are studied as well: existence of a time interval when clique or quasi-clique can exist. Numerical examples illustrate the approach and problems.

## I. INTRODUCTION

In many contemporary complex systems, the significance of function system testing is increasing (e.g., [7], [24]). In two recent decades, a central role of multi-function system testing has been pointed out in many studies (e.g., [9], [24], [28]), i.e., combinations of system functions can lead to general system faults. This situation is based on integration (fusion) of local faults into a general system faults. Here a local fault corresponds to a system unit/components and each local fault is a result of a certain system function. Often coordination of concurrent system functions can be organized at a insufficient level. In many recent well-known system faults (e.g., power stations, offshore drilling platforms, airplane crashes), coordination among concurrent system functions was not OK (by computer systems, by maintenance documentations, etc.). In contemporary distributed complex systems, there exist many different users (they are not often coordinated), many different support teams (i.e., maintenance, management; they can be coordinated at insufficient level), and many different processes (they can be coordinated at insufficient level). Evidently, this kinds of applied situations for complex distributed systems

Mark Sh. Levin: Inst. for Inform. Transmission Problems, Russian Acad. of Sci., Moscow 127994, Russia. Email: mslevin@acm.org.

have to be examined via various approaches. Thus, a system analysis of combinations for concurrent system functions is a crucial direction in system testing (*e.g.*, [9], [24], [28]).

In our paper, the above-mentioned type of applied situations for complex distributed systems is considered and described (i.e., problem, models, solving scheme). Our framework is based on clique-based fusion of graph streams for multi-function system testing ([22], [23], [24]).

Now let us consider a simplified example for a modular system consisting of the following components: basic facility $s_1$, control subsystem $s_2$, safety subsystem $s_3$, utilization personnel $s_4$, maintenance/ testing personnel $s_5$, and personnel for remote control $s_6$. Three functions are examined (Fig. 1): utilization function $f_1 : \{s_1, s_2, s_3, s_4\}$, maintenance/ testing function $f_2 : \{s_1, s_2, s_3, s_5\}$, remote control function $f_3 : \{s_1, s_2, s_6\}$. In the case when a coordination between the above-mentioned concurrent functions is wrong, the following situation can be met: (i) basic facility $s_1$ is out of service (by utilization personnel $s_4$, i.e., via function $f_1$); (ii) safety subsystem $s_3$ is out of service (by an action of maintenance/testing personnel $s_5$, i.e., via function $f_2$); and (iii) control subsystem $s_2$ is under a wrong mode (by a wrong action of personnel for remote control $s_6$, i.e., via function $f_3$). Fig. 1 depicts the examined situation (ordinal estimates of graph vertices are shown in circles; the following estimates for nodes are used: 1 corresponds to a wrong mode, 2 corresponds to about 'OK', 3 corresponds to 'OK').

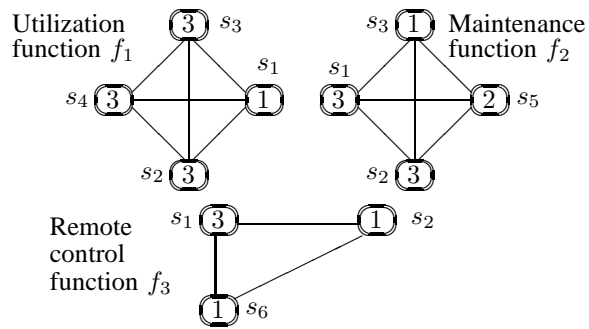

Fig. 1. Illustration for simplified example

As a result, a combination of interconnected system components (as clique) $\{s_1, s_2, s_3\}$ will lead to a combined system fault.

Our approach to system testing is based on a layered scheme (Fig. 2, adopted from [23] and [24]). Generally, the system testing framework consists of several stages as follows [22]:

*Stage 1.* Unit/component testing (e.g., [18]).

*Stage 2.* Analysis of system functions, their interconnections, and design of function clusters (functions which are executed jointly/concurrently, i.e., at the same time moment) ([23], [24]).

*Stage 3.* Design of an integrated graph over system components for each function cluster.

*Stage 4.* Revelation of clique (or quasi-clique) in the integrated graph (e.g., [29], [30], [31]).

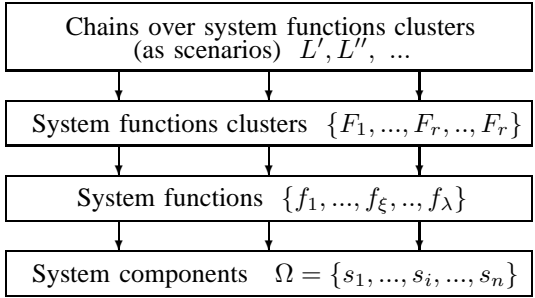

Fig. 2. Layered system testing scheme

A simplified framework of the testing process is depicted in Fig. 3 (an integrated graph corresponds to function cluster $F' = \{f_1, ..., f_\xi, ..., f_\lambda\}$). Revelation of clique in the integrated graph is considered as a fusion problem or structure mining (e.g., [8], [15]). Thus, our framework is based on revelation (mining) of cross-graphs cliques or quasi-cliques ([24], [31]). Many well-known traditional methods may be used for the analysis and revelation of cliques (or quasi-clique) in graphs (e.g., [1], [12], [16], [29], [30], [31]).

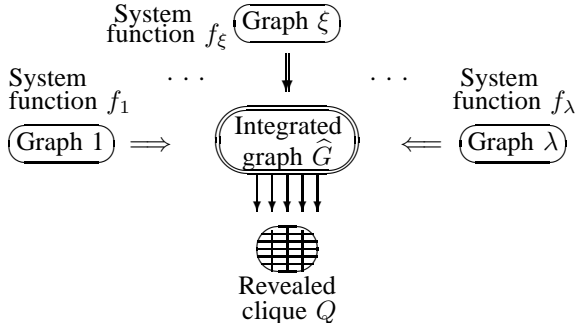

Fig. 3. Revelation of clique (quasi-clique)

## II. PRELIMINARIES

Here the basic sets are described [24]. Let $\Omega = \{s_1, ..., s_i, ..., s_n\}$ be a set of subsystems or main system components. Let $f = \{f_1, ..., f_\xi, ..., f_\lambda\}$ be a set of system functions. For each system function $f_\xi$ there is the following: (i) a subset of $\Omega(f_\xi) \subseteq \Omega$ that consists of components which are used for system function $f_\xi$ and (ii) a graph (usually a complete graph) over elements of $\Omega(f_\xi)$ as $G(f_\xi) = (\Omega(f_\xi), E(f_\xi))$.

There is a set of system function clusters $F = \{F_1, ..., F_r, ..., F_p\}$ where $F_r \subseteq f$; each system function cluster $F_r$ ($r = \overline{1,p}$) is a subset of system function set $F$ and for each system function cluster its elements (i.e., corresponding functions) are executed together at the same time moment. For each system function cluster $F_r$ it is reasonable to examine the corresponding integrated graph as follows: $G(F_r) = G(\Omega(F_r), E(F_r)) = \cup_{f_j \in F_r} G(f_j)$, where $\Omega(F_r) = \cup_{f_j \in F_r} \Omega(f_j)$ and $E(F_r) = \cup_{f_j \in F_r} E(f_j)$.

For each subsystem $s_i$ a system test procedure (as unit test) is used and the procedures lead to ordinal results (i.e., states, colors) for each system component, e.g., $[1, 2, 3, 4]$ (where 1 corresponds to "out of service", 2 corresponds to "major faults", 3 corresponds to "minor faults", 4 corresponds to "trouble free service"). As a result, the graphs with ordinal weights of vertices (or colored graphs) are obtained. Finally, for each function cluster $F_r$ we can examine the corresponding colored integrated graph $\widehat{G}(F_r) = \widehat{G}(S(F_r), E(F_r))$.

In more complicated situation, the unit test results can be different for different system functions. Then the integration process is based on the following rule: in the case of difference of colors for the same vertex of graphs for different system functions (e.g., $f_{j_1}$ and $f_{j_2}$) the 'worst' color is selected.

Let $Q_h$ be a clique over $h$ vertices (e.g., $Q_4$, here the estimate of each clique vertex is: $\leq l$, $l = 1, 2, 3, 4$), and "quasiness" or "approximation" (by number of vertices or by estimates of vertices) will be denoted by "widetilde" (e.g., $\widetilde{Q}_4$). Thus, in the integrated colored graph $\widehat{G}(F_r)$ the following kinds of substructures (subgraphs) are under examination (by a rule: in the structure each vertex color $"= 1"$, $"\leq 2"$, etc.). Fig. 4 illustrates 4-vertex structure (ordinal estimates of graph vertices are pointed out in circles):

1. Clique, dimension of the clique equals (or more than) the number of functions in $F_r$ (Fig. 4a): $Q_r$.

2. Quasi-clique by edges/interconnection (an edge is absent) (Fig. 4b): $\Phi_r$.

3. Quasi-clique by vertices (in the revealed subgraph not all vertices have estimate $\leq l$, $l = 1, 2, ...$) (Fig. 4c): $\Theta_r$.

4. Quasi-clique by vertices and by edges (Fig. 4d): $\widetilde{\Theta}_r$ or $\widetilde{\Phi}_r$.

5. Sub-clique or clique with less dimension (i.e, the number of vertices in the clique $v$ is less than the number of functions in $F_r$ (Fig. 4e): $Q_v$, $v < p$).

6. Quasi sub-clique (structural approximation): (i) by vertices, (ii) by edges, (iii) by vertices and edges ($\widetilde{Q}$, etc.).

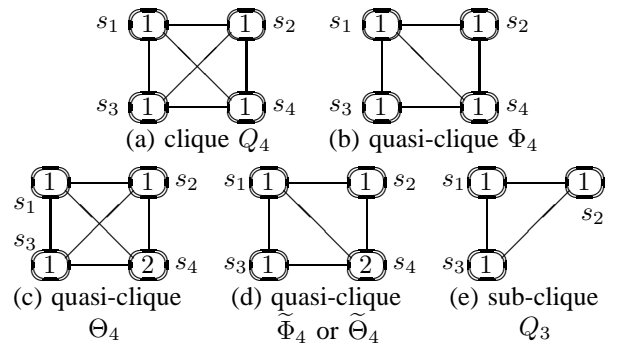

Fig. 4. Examples of clique and quasi-cliques

It is reasonable to point out the following:

**I.** A situation when the estimate of system components equals 1 (i.e., "out of service") is crucial and can lead to a system fault. This kind of situation is a "traditional" one in system testing.

**II.** A situation when several interconnected (by time and/or system work) system components have estimates at a "medium level" (e.g., "major fault" or "minor faults") can lead to a system fault in complex systems (e.g., [9]). Thus, our main efforts in this paper are targeted to this kind of a system situation (Fig. 5, here estimates of vertices are: $\leq 3$). On the other hand, the further examination can be mainly based for estimates of system components at the levels 1 and 2 because after a shift of the ordinal evaluation scale (i.e., [1,4]) this kind of mathematical problems will be obtained.

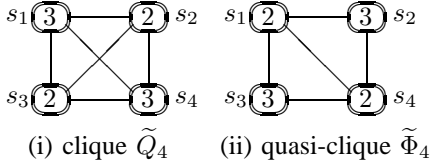

Fig. 5. Examples of system faults

### III. BASIC PROBLEM

Here a basic problem (*problem 1*) is described. Structural fusion of quality estimates for system units/components upon ordinal scale is illustrated in Fig. 6, Fig. 7, and Fig. 8. The estimates for system components are depicted in Fig. 6 (components $s_3$, $s_5$, and $s_6$: 1; component $s_1$: 2; and component $s_7$: 3; other components: 4). The basic problem (*Problem 1*) is:

*Find for multi-function situation (in the integrated graph for function cluster $F_r$) clique $Q_h$ (number of nodes equals the number of functions in cluster $F_r$ or more with the estimate level $\leq l$ ($l = 1, 2, ...$).*

Thus, the well-known clique problem is considered:

*Find the largest clique (complete subgraph) $Q$ in an undirected graph.*

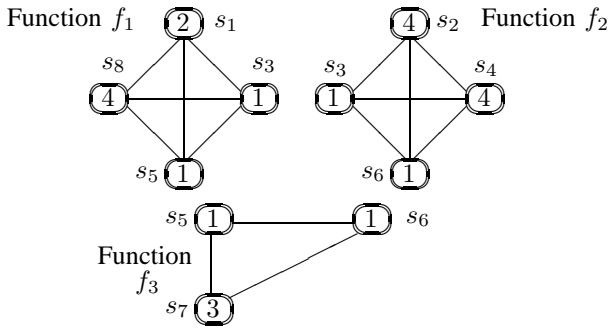

Fig. 6. System structure and estimates

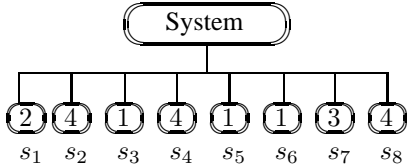

Fig. 7. Cluster of system functions

The obtained set of system vertices/units has to be examined as "critical unit subset" or "system syndrome" (analogue of "syndrome" in medicine). In Fig. 6 and Fig. 7, $F_r = \{f_1, f_2, f_3\}$. Further, in Fig. 8 the clique vertices are the following: $\{s_3, s_5, s_6\}$. Here estimates of the vertices above equal 1.

Clearly, the clique $Q_\lambda$ may be absent. In this case it may be important to search for quasi-clique $\widetilde{Q}_\lambda$ or cliques which contain less number of vertices.

Thus, another problem of structural fusion is (*problem 1a*):

*Find quasi-clique $\widetilde{Q}_\lambda$ for the multi-function situation, i.e., without some interconnection/edges or/and with the estimate level: $\leq l$, ($1 < l \leq 3$).*

Examples of quasi-cliques (vertex sets) are the following (Fig. 8): (a) $\{s_1, s_3, s_5, s_6\}$, estimates: $\leq 2$; (b) $\{s_1, s_3, s_5, s_6, s_7\}$, estimates: $\leq 3$. Note, it may be often possible (and reasonable) to reveal several cliques (or quasi-cliques).

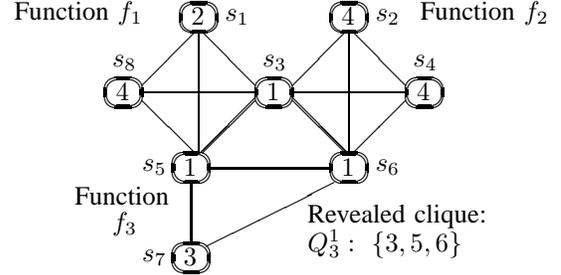

Fig. 8. Integrated graph for function cluster

### IV. ADDITIONAL PROBLEMS

#### A. Basic Sequences/Streams

In the case of graph streams, an illustration is depicted in Fig. 9.

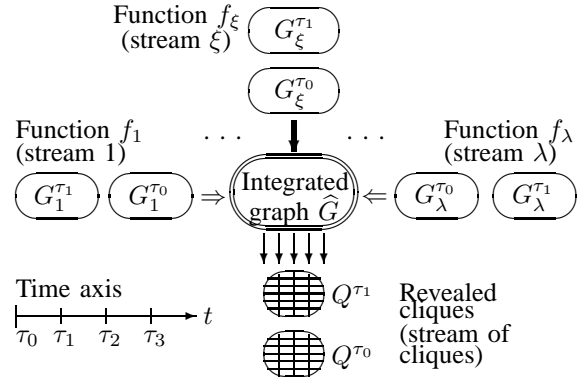

Fig. 9. Revelation of clique over graph streams

Here a time axis is considered as follows:
$t = \{\tau_0, \tau_1, \tau_2, \tau_3, \tau_4, \tau_5, ...\}$.
As a result, the following is examined:
(i) sequence of states for system components ($j = \overline{1, k}$):
$s_j(t) = \{s_j^{\tau_0}, s_j^{\tau_1}, s_j^{\tau_2}, s_j^{\tau_3}, s_j^{\tau_4}, s_j^{\tau_5}, ...\}$,
where $s_j(\tau_\eta) \in \{0, 1\}$ ($\eta = 0, 1, 2, 3, 4, 5, ...$), $s_j(\tau_\eta) = 1$ if function $j$ is used at time $\tau_\eta$ and $s_j(\tau_\eta) = 0$ otherwise;
(ii) sequence of system functions ($\xi = \overline{1, \lambda}$):
$f_\xi(t) = \{f_\xi^{\tau_0}, f_\xi^{\tau_1}, f_\xi^{\tau_2}, f_\xi^{\tau_3}, f_\xi^{\tau_4}, f_\xi^{\tau_5}, ...\}$,
where $s_j(\tau_\eta) \in \{0, 1\}$, $s_j(\tau_\eta) = 1$ if function $j$ is used at time $\tau_\eta$ and $s_j(\tau_\eta) = 0$ otherwise);
(iii) sequence of graphs for system functions ($\xi = \overline{1, \lambda}$):
$G_{f_\xi}(t) = \{G_{f_\xi}^{\tau_0}, G_{f_\xi}^{\tau_1}, G_{f_\xi}^{\tau_2}, G_{f_\xi}^{\tau_3}, G_{f_\xi}^{\tau_4}, G_{f_\xi}^{\tau_5}, ...\}$
(if $s_j^{\tau_\eta} = 0$ the corresponding graph $G_{f_\xi}^{\tau_\eta}$ is empty);

(iv) sequence of graphs for system function cluster ($r = \overline{1,p}$):

$G_{F_r}(t) = \{G_{F_r}^{\tau_0}, G_{F_r}^{\tau_1}, G_{F_r}^{\tau_2}, G_{F_r}^{\tau_3}, G_{F_r}^{\tau_4}, G_{F_r}^{\tau_5}, ...\}$

(if $s_j^{\tau_\eta} = 0$ the corresponding graph $G_{F_r}^{\tau_\eta}$ is empty).

Here a chain of system function clusters (e.g., $L = < F', F'', F''' >$) is considered as a scenario (in general, scenario can have a more complicated type, e.g., tree, network).

### B. Other Problems

The set of additional problems involves the following:

*Problem 2.* Revelation of clique when the number of vertices is less than the number of functions in the function cluster (i.e., sub-clique).

*Problem 3.* Dynamical problems (vertex colors are functions of time): *3.1.* existence of a time interval where clique exists; *3.2.* existence of a time interval where quasi-clique exists.

*Problem 4.* Analysis of time intervals when clique (or quasi-clique) exists and maintenance of the clique (quasi-clique) as some "critical" structure (substructure). As a result, a track for a special structure (e.g., clique $Q$) can be obtained: $T_Q$.

*Problem 5.* Design of actions as composite plans to destroy the critical substructure(s) (i.e., clique(s), quasi-clique(s)).

## V. EXAMPLE

Let us consider a numerical example. Table 1 and Table 2 contain a description of the examined sets of system functions and function clusters: $\{f_1, f_2, f_3, f_4, f_5\}$ and $\{F_1, F_2, F_3\}$.

Table 1. System functions

| System function | System components | | | | | | | |
|---|---|---|---|---|---|---|---|---|
| | $s_1$ | $s_2$ | $s_3$ | $s_4$ | $s_5$ | $s_6$ | $s_7$ | $s_8$ |
| $f_1$ | ⋆ | | ⋆ | | ⋆ | | | ⋆ |
| $f_2$ | | ⋆ | ⋆ | ⋆ | | ⋆ | | |
| $f_3$ | | | | | ⋆ | ⋆ | ⋆ | |
| $f_4$ | ⋆ | | | | ⋆ | | | |
| $f_5$ | | | | | ⋆ | ⋆ | | ⋆ |

Table 2. System function clusters

| System function clusters | System functions | | | | |
|---|---|---|---|---|---|
| | $f_1$ | $f_2$ | $f_3$ | $f_4$ | $f_5$ |
| $F_1$ | ⋆ | ⋆ | ⋆ | | |
| $F_2$ | | ⋆ | | ⋆ | ⋆ |
| $F_3$ | | | ⋆ | | ⋆ |

The following time axis is considered: $\{\tau_0, \tau_1, \tau_2, \tau_3, \tau_4, \tau_5\}$ (i.e., $\tau_\eta$, $\eta = \overline{0,5}$). The function cluster chain (a scenario) is as follows: $L' = < F_2^{\tau_0}, F_1^{\tau_1}, F_3^{\tau_2}, F_1^{\tau_3}, F_1^{\tau_4}, F_2^{\tau_5} >$ where *upper index* corresponds to time moment. Fig. 8 depicts integrated graph $\widehat{G}(F_1)$, Fig. 10 depicts two integrated graphs: $\widehat{G}(F_2)$ and $\widehat{G}(F_3)$.

Further, the following basic problems are under examination: problem 3, problem 4, problem 5. Fig. 11 depicts state streams for system components (vertices) $s_1$, $s_2$, $s_3$, $s_4$, $s_5$, $s_5$, $s_6$, $s_7$, and $s_8$ while taking into account time axis above. In addition, Fig. 11 contains the following: (i) integrated graphs $\{\widehat{G}\}$, (ii) revealed structures (here: $Q_3$), and (iii) obtained tracks of the revealed structures (here: $T_{Q_3}$).

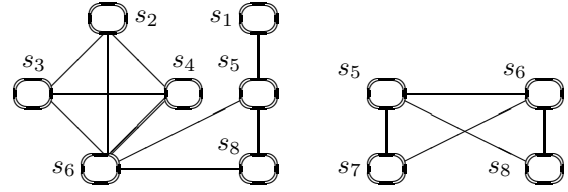

(a) integrated graph $\widehat{G}(F_2)$  (b) integrated graph $\widehat{G}(F_3)$

Fig. 10. Integrated graphs for function clusters

Now the following solutions can be pointed out:

*Problem 3.1.* Clique $Q_3$ : $\{s_3, s_5, s_6\}$ can be revealed for time: $\tau_1, \tau_3, \tau_4$ (estimates of vertices equal 1).

*Problem 4.* For time interval $[\tau_1, \tau_5]$ there exists a structure with vertices: $\{3, 5, 6\}$ while taking into account a well-known engineering "track initiation rule: $2\ from\ 3$" (i.e., for time interval with length 3 clique is revealed 2 times). As a result, it is reasonable to initiate at time moment $\tau_3$ clique $Q_3$ : $\{s_3, s_5, s_6\}$ (estimates of vertices equal 1). Note, the rule above (the rules of these kind) is used as "track maintenance rule" as well. After that it is possible to maintain this structure (by the rule above) as an event (i.e., to check the initiation rule above at each discrete time moment, for example: $\tau_4$, $\tau_5$ ).

*Problem 5.* Clearly, to destroy the event above (i.e., $T_{Q_3}^{\tau_3}$, $T_{Q_3}^{\tau_4}$, $T_{Q_3}^{\tau_5}$) it is necessary to destroy clique $Q_3$ at time moment $\tau_3$ by improvement of state for $s_5$ (or $s_6$) (i.e., improvement of the estimate: $1 \to 2$ or $1 \to 3$ or $1 \to 4$).

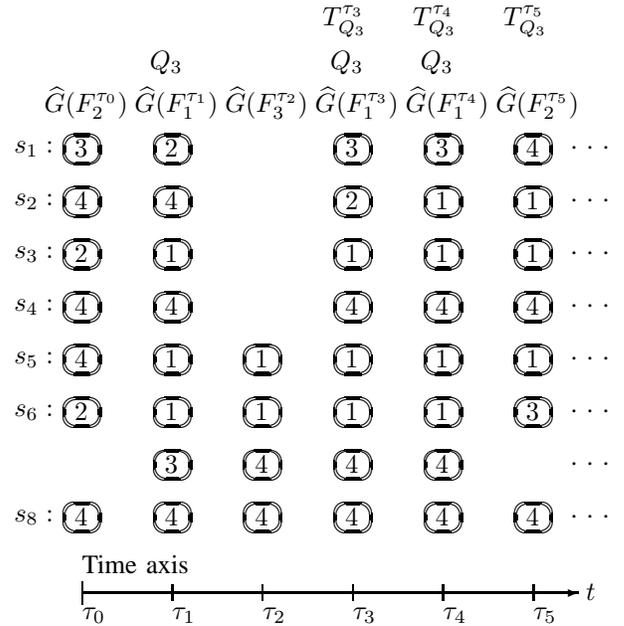

Fig. 11. Example of graph streams

## VI. GENERALIZED TESTING SCHEME

A generalized testing scheme is presented in Fig. 12. Evidently, the generation of test inputs as sequences $\{s_j(t), j = \overline{1,k}\}$ can be based on two methods: (a) previous practice,

(b) special simulation approaches (e.g., Monte Carlo or quasi Monte Carlo methods [33]).

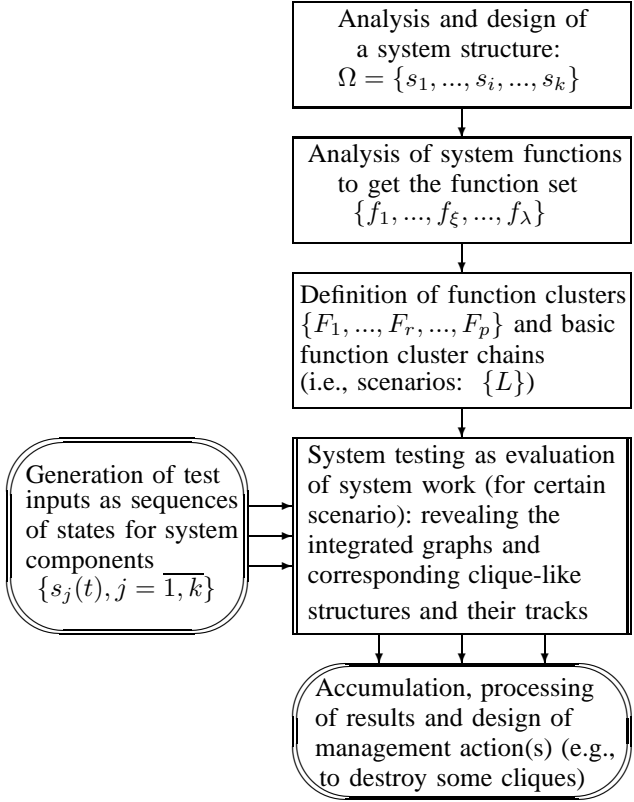

Fig. 12. Generalized testing scheme

The suggested system testing framework can be used for system maintenance as well (Fig. 13): (1) evaluation of system components $\{s_j\}$; (2) revelation of the executed system function clusters (i.e., $F_r$); (3) revelation of integrated graphs, corresponding "critical" clique-like structures, and the structure tracks; and (4) analysis and planning of system maintenance: (a) prediction: for system components and for whole system, (b) design of maintenance plan (e.g., improvement actions for system components).

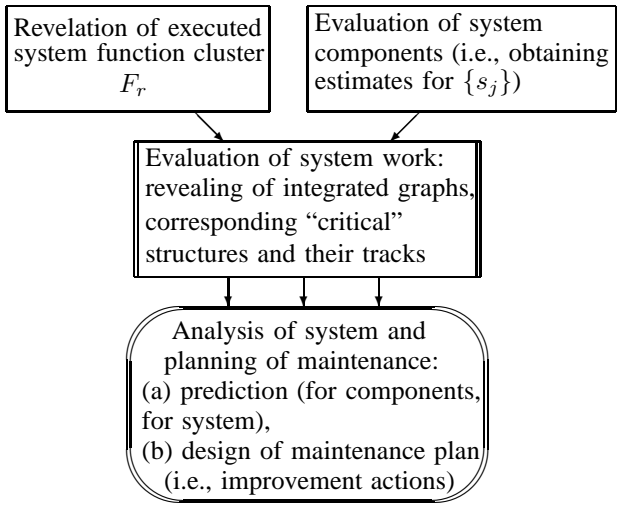

Fig. 13. Scheme of system maintenance

## VII. Conclusion

In the paper, a new system framework for multi-function system testing has been suggested. The positive property of the framework consists in the following: concurrent system functions can be analyzed. The system framework is a basis for modification and adaptation.

In general, it is reasonable to point out the significance of data stream systems (e.g., [2], [8], [32], [34]) which are widely used in many domains (data/knowledge summarization, image processing, system reliability analysis, initiation of target tracks in sensor systems, etc.). The key parameters for the systems above involve the following: (1) number of streams (one, many), (2) type of data, e.g., values (binary, ordinal, continuous), structures (i.e., preferences, graphs), (3) size of time window (i.e., number of series time moments which are jointly analyzed). A simplified typology of the systems may be considered as follows (e.g., [2], [13], [20], [25], [27], [32], [34]):

(a) static case for $m$ streams: (i) summarization of values (binary, ordinal, continuous); processing methods: histograms, rule "$k$ of $m$", diagnosis techniques (e.g., closeness to centers of specified clusters) (e.g., [2]); (ii) aggregation of structures (e.g., [10], [14], [17], [19], [20], [21], [26]); processing methods (e.g., decision making, knowledge engineering, image processing): building a maximum substructure (e.g., consensus, median), building a minimum superstructure;

(b) dynamic case (one stream, window for $n$ time moments); processing methods, for example: rule "$k$ of $m$" (e.g., an engineering technique for initiation/maintenance of target tracks, analysis of system reliability, fusion of image sequences, revelation of patterns from time series of graphs) (e.g., [3], [8], [32]); and

(c) combined dynamic case ($m$ streams and window for $n$ time moments); composite processing methods (e.g., dynamic decision making based on Markov decision processes or dynamic decision networks) (e.g., [3], [4], [5], [6], [11]).

Three possible evident strategies can be used in case (c) above:

*strategy 1*: (i) integration of data for each stream by a time window (case b), (ii) summarization of results for $m$ streams (case a);

*strategy 2*: (i) summarization of data of $m$ streams at each time moment (case a); (ii) integration of results via a time window (case b); and

*strategy 3*: combined scheme.

Our suggested framework implements *strategy 2* above: (i) fusion of graphs at each time moment with revelation of clique (or quasi-clique), and (ii) usage of rule "$k$ of $m$".

Note, our material has only a preliminary character as the first step and it is targeted to problem(s) formulation and solving scheme description (i.e., a new system framework). Next research efforts could include various investigations, for example: *1.* study of the problems with some weights of system components or/and their faults; *2.* exploration of cliques (quasi-cliques) as special kinds of composite events (as "generalized system syndromes") (to generate a set of possible composite systems faults); *3.* study of stability issues for considered solving schemes; and *4.* design of a special simulation computer environment based on the suggested framework.

**Mark Sh. Levin** received the M.S. degree in Radio Engineering from Moscow Techn. Univ. for Communication and Informatics (1970), the M.S. degree in Mathematics from Lomonosov Moscow State Univ. (1975), the Ph.D. degree in Systems Analysis from Russian Acad. of Sci. (1982). He conducted his research projects in Russia, Israel, Japan, and Canada. Now Dr. Levin is a Senior Research Scientist in Inst. for Information Transmission Problems of Russian Acad. of Sci. His research interests include systems engineering, system design, combinatorial optimization, decision making, networking, and education in engineering, computer science, and applied mathematics. Dr. Levin has authored many research articles and three books. He is a member of IEEE, ACM(SM'06), SIAM, INCOSE. More information can be found at http://www.mslevin.iitp.ru/